\documentclass[showpacs,preprintnumbers,amsmath,amssymb,aps,twocolumn]{revtex4}
\usepackage{graphicx}% Include figure files
\usepackage{bm}% bold math
\usepackage{times}

\begin{document}
\title{Nuclear quadrupole resonances in compact vapor cells: 
			\\the crossover between the NMR and the nuclear quadrupole resonance interaction regimes}
\author{E.A. Donley}
\author{J.L. Long}
\author{T.C. Liebisch}
\author{E.R. Hodby}
\author{T.A. Fisher}
\author{J. Kitching} 
\affiliation{%
NIST Time and Frequency Division\\
325 Broadway, Boulder, CO 80305}%
\date{\today}
\begin{abstract}
We present an experimental study that maps the transformation of nuclear
quadrupole resonances from the pure nuclear quadrupole regime to the 
quadrupole-perturbed Zeeman regime. The transformation presents an interesting 
quantum-mechanical problem, since the quantization axis changes from being 
aligned along the axis of the electric-field gradient tensor to being aligned 
along the magnetic field. The large nuclear quadrupole shifts present in our system
enable us to study this regime with relatively high resolution. 
We achieve large nuclear quadrupole shifts for 
$I = 3/2$ $^{131}$Xe by using a cube-shaped 1~mm$^3$ vapor cell with walls of 
different materials. The enhancement of the NQR shift from the cell wall materials 
is a new observation that opens up an additional adjustable parameter to tune and 
enhance the nuclear quadrupole interactions in vapor cells. As a confirmation
that the interesting and complex spectra that we observe are indeed expected, 
we compare our data to numerical calculations and find excellent 
agreement. 
\end{abstract}

\maketitle
\section{Introduction}
Any atom that has a nuclear spin $I \ge 1$ has a nuclear electric quadrupole 
moment, whose interactions with electric field gradients can cause shifts of 
the nuclear magnetic energy levels. There is a large body of literature on 
the interactions of nuclear quadrupole
moments with electric field gradients. As far back as the 1950s, studies were
performed in crystals both in the regime where the nuclear quadrupole 
interaction caused weak perturbations to the nuclear magnetic resonance (NMR)
spectra \cite{Pound1950}, as well as in the pure nuclear quadrupole resonance
(NQR) regime, where little or no Zeeman interaction was present 
\cite{Bloom1955}. Solutions for the transition energies between nuclear spin 
sublevels were found for both regimes by use of perturbation theory (for a 
review, see \cite{Abragam1961}) by aligning the quantization axis along the 
principal axis of the electric field gradient tensor in the NQR regime and 
along the axis of the magnetic field in the NMR regime. As with these first 
experiments, most NQR studies have either been distinctly in either the NQR 
or the NMR regimes. To our knowledge, prior to our work reported here, the 
transformation of the NQR spectra from the NQR to the NMR regimes had not been 
observed experimentally.

Cohen-Tannoudji first suggested that in addition to arising from electric 
field gradients from ionic bonds in a crystal, quadrupolar coupling could 
occur between nuclei and electrical field gradients present at 
the nucleus during wall collisions for atoms in vapor cells 
\cite{Cohen-Tannoudji1963}. Since then, nuclear quadrupole resonances in 
vapor cells have been studied for many systems including $I=3/2$ $^{201}$Hg 
\cite{Simpson1978, Heimann1981A, Heimann1981B, Lamoreaux1986, Lamoreaux1989}, 
$I=9/2$ $^{83}$Kr \cite{Volk1979}, $I=3/2$ $^{131}$Xe 
\cite{Kwon1981, Wu1987, Wu1988, Wu1990, Butscher1994, Appelt1994}, 
and $I=3/2$ $^{21}$Ne \cite{Chupp1990}. Much of this work has been of 
basic interest from a fundamental physics standpoint 
\cite{Simpson1978, Heimann1981A, Heimann1981B, Volk1979, Kwon1981, Wu1987, 
Wu1988, Wu1990, Butscher1994, Appelt1994} and for tests of fundamental 
symmetries \cite{Lamoreaux1986, Lamoreaux1989, Chupp1990, Majumder1990}. 
There have also 
been proposals for using these systems for the practical application of 
rotation sensing \cite{Simpson1964, Grover1979}. Changes to the NQR shifts 
in the crossover regime could lead to systematic errors in precision 
measurements and offsets in rotation sensors.

For much of the NQR work that has been performed in vapor cells, the NQR lines 
were not clearly resolved and the NQR splitting caused a slow beating or a 
nonexponential decay of the nuclear polarization \cite{Simpson1978, Volk1979, 
Kwon1981, Heimann1981A, Heimann1981B}. The beat frequencies depended on the 
orientation of the cell symmetry axis in the magnetic field, and went to zero 
at the magic angle of $54.7\,^{\circ}$ \cite{Simpson1978}, which indicated that 
the axis of symmetry of the electric field gradient was aligned along the cell 
symmetry axis. 

In a series of papers from Happer's group at Princeton 
\cite{Wu1987,Wu1988,Wu1990}, Wu et al. saw much stronger interactions such that 
the NQR lines could be clearly resolved by using highly
asymmetric cells. They performed a detailed perturbation-theory solution for 
the NQR shift in the NMR regime, accounting for pressure-dependent diffusion 
and cell shape, and formulated the results to give a microscopic description 
of the interaction \cite{Wu1988}. Ignoring complications from diffusion, they 
expressed the NQR shifts for the 
$\left|-3/2\right\rangle \left\langle -1/2 \right|$ and 
$\left|1/2\right\rangle \left\langle 3/2 \right|$ coherences as
\begin{equation}
\Delta\Omega = \pm \frac{v S}{2 V} \frac{1}{2I-1} \int_S \frac{dS'}{S} 
\left\langle \theta \right\rangle \left[\frac{3}{2}\cos^2{\psi}-\frac{1}{2} \right]  ~, 
\label{Eq1}
\end{equation}
which is an integration of the nuclear quadrupole interaction over the cell 
walls. Here $v$ is the atom velocity,  $\left\langle \theta \right\rangle$ is 
the mean twist angle per wall adhesion,  $S$ is the cell surface area, $V$ is 
the cell volume, $I$ is the nuclear spin, and  $\psi$ is the angle between the 
local surface normal (directed out of the cell) and the magnetic field. Here 
we put $\left\langle\theta\right\rangle$  within the integral to allow for the 
possibility that the cell walls are of different materials. Integrating Eq. 
\ref{Eq1} for a cylindrical cell gives
$\Delta\Omega =  \pm \Delta\Omega_0 \rm{P}_2\left(\cos{\varphi}\right)$,
where $\varphi$ is the angle between the cell symmetry axis and the direction 
of the magnetic field and 
$\rm{P}_2\left(x \right)=\frac{1}{2}\left(3 x^2 -1 \right)$. 
$\Delta\Omega_0 = v A \left\langle \theta \right\rangle S/4V$ is proportional 
to the atom velocity, the surface to volume ratio of the cell, and an 
asymmetry parameter, $A$, which goes to zero when the cell height and diameter 
are equal.
Wu et al. verified their theory with detailed experiments and determined 
$\left\langle \theta \right\rangle=38\left(4\right)\times 10^{-6} \,\rm{rad}$ 
for pyrex \cite{Wu1990}. Experiments later performed by Butscher et al. 
\cite{Butscher1994} revealed similar behavior. 

All of the experiments described so far were performed at magnetic fields high 
enough that the nuclear quadrupole interaction was well described as a 
perturbation to the magnetic Larmor resonances. Appelt et al. performed 
studies on $^{131}$Xe in the limit of zero applied magnetic field and measured 
deviations from Berry's adiabatic phase under rotation \cite{Appelt1994}. They 
solved the $I=3/2$ Hamiltonian by including terms for the NQR interaction and 
spatial rotations and showed that mixing of the nuclear spin sublevels through 
rotation makes all six transitions between nuclear sublevels allowed 
($\Delta m$ = 1, 2, and 3).

All of this work was either distinctly in the NMR 
\cite{Volk1979, Kwon1981, Wu1987, Wu1988, Wu1990, Butscher1994} or the NQR 
\cite{Appelt1994} regime. Here we bridged the two regimes by continuously 
tracing the transformation from the pure NQR regime to the 
quadrupole-perturbed Zeeman regime. We achieved a large NQR splitting not by 
using geometrically asymmetric cells as Wu et al. \cite{Wu1987} did , but by 
using a 1~mm$^3$ cubic cell with walls of different materials. A cubic charge 
distribution would not ordinarily cause an NQR splitting \cite{Heimann1981B}, 
but the microscopic 
surface interactions with the different wall materials lower the symmetry of 
the system. The small cell size  enhances the NQR splitting because the shifts 
are proportional to the surface to volume ratio. 
The enhancement of the NQR shift from the cell wall materials is a new 
observation that gives one an additional adjustable parameter to tune the 
electric field gradient and the NQR interaction in vapor cells.
As a confirmation that the complex
spectra that we observe are expected, we compare the transformation 
of the resonance lines to theoretical models and find excellent agreement.

\section{Experiment}
\subsection{Techniques and Apparatus}
\begin{figure}[ht]
\includegraphics[width=8.0cm]{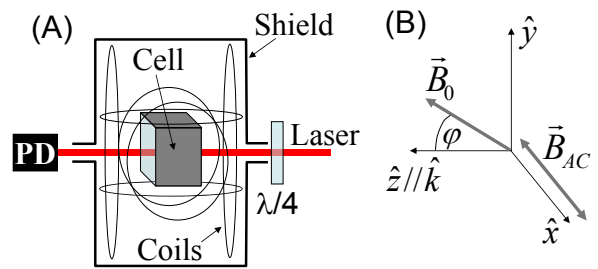}
\caption{\label{Apparatus}(A). The apparatus. The 1~mm$^3$ cell is roughly 
centered on an orthogonal three-axis set of magnetic coils. A laser beam of 
maximum power $3 \,$mW enters the cell. The transmitted power is detected by a 
photodiode.  (B). The coordinate system: the $\hat{z}$-axis coincides with the 
light propagation direction, $\hat{k}$. The magnetic field direction is in the 
$\hat{y}$-$\hat{z}$ plane rotated by an angle $\varphi$ from the 
$\hat{z}$-axis. }
\end{figure}

Fig. \ref{Apparatus}A is a schematic drawing of our apparatus. The 
microfabricated sample cell of volume 1~mm$^{3}$ was etched in silicon and 
sealed with pyrex \cite{Knappe2005}. The cell contains $^{87}$Rb, buffer gases 
of N$_2$ and Ne at 10 and 600 torr, respectively, and 10 torr of Xe gas at 
natural abundance. Xe has 
two active NMR isotopes: spin-1/2 $^{129}$Xe (26.4\% abundance) and spin-3/2 
$^{131}$Xe (21.2\% abundance). The cell is cubic and has four silicon walls and 
two pyrex windows. The cell is heated to $145~^{\rm{o}}$C and mounted at the 
center of a set of three orthogonal magnetic coils. The coils are surrounded 
by a four-layer magnetic shield (one layer shown in Fig. \ref{Apparatus}A). A 
circularly polarized laser beam optically pumps and probes the Rb atoms 
through the pyrex cell windows along the $\hat{z}$-axis. The Rb polarizes the 
Xe atoms through spin-exchange optical pumping \cite{Walker1997}. The Rb  also 
functions as a magnetometer and is used to sense the magnetic fields generated 
by the Xe atoms \cite{Grover1978, Keiser1977}. 

We use a field switch technique to initiate precession of the Xe atoms. During 
the pump phase, the Xe polarization builds and reaches steady state. At the 
start of the probe phase, a DC magnetic field, $B_0$, is turned on in the 
$\hat{y}$--$\hat{z}$ plane and the Xe atoms start to precess. The angle of 
$\vec{B}_0$ with respect to the $\hat{z}$-axis (the cell symmetry axis), 
$\varphi$, is varied, depending on the experiment. An AC magnetic field, 
$\vec{B}_{AC}$, of rms amplitude $\sim 1 \mu$T and frequency $\sim 2\,$kHz 
drives the Rb atoms and is applied along the $\hat{x}$-axis. This AC drive 
also references a lock-in amplifier that measures the modulation of the 
transmitted power at the Rb drive frequency. The applied field geometry is 
shown in Fig. \ref{Apparatus}B. We are also able to observe signals in many 
other field configurations, but we have found this configuration to give the 
best signal-to-noise ratio. Our geometry for pumping and probing is very 
similar to the technique used by Volk et al. \cite{Volk1980}.

After a field switch, a free induction decay (FID) signal is observed at the 
output of the lock-in amplifier. Fig. \ref{SampleData} shows an FID signal and 
its Fourier transform. In most cases, we used an acquisition time of $32.8\,$s 
on our spectrum analyzer, which gave a frequency resolution of $30.5\,$mHz. 
Since the acquisition time was much longer than our typical T$_2$ time of 
$\sim 5\,$s, we compromised on signal-to-noise ratio to achieve higher 
frequency resolution.

One factor that complicates our estimating the field magnitude and angle is 
the relatively large field generated by the Rb atoms as sensed by the Xe atoms 
\cite{Schaefer1989}, the magnitude of which is
$B_{Rb} = \mu_B \frac{8\pi\kappa}{3} \frac{\mu_0}{4\pi} n_{Rb} P$.
$\mu_B$ is the Bohr magneton,  $\kappa =730$ is the hyperfine contact 
enhancement factor \cite{Walker1989},  $\mu_0$ is the magnetic constant, 
$n_{Rb}$ is the Rb density, and $P$ is the Rb polarization. At our laser 
intensity of 300 mW/cm$^{2}$ and temperature of $145~^{\rm{o}}$C, we measure 
$B_{Rb} = 200 \,$nT, which corresponds to $P = 30\,\%$.

To simplify controlling the total field in the presence of this large offset 
field, for most of our measurements we divide our applied field into two 
components such that our total field is 
$\vec{B}_{tot} = \vec{B}_0 + \vec{B}_{Rb} + \vec{B}_{comp}$. We set the compensation 
field, $\vec{B}_{comp}$, equal to $-\vec{B}_{Rb}$, such that we can determine our 
total field as sensed by Xe from the variable component $\vec{B}_0$ alone.
Offsetting $\vec{B}_{Rb}$ is made easier by having a high pumping rate so that 
$\vec{B}_{Rb} \parallel \hat{k}$ is independent of the angle of $\vec{B}_0$. 
Then $\vec{B}_{Rb}$ can be nulled out with a constant field parallel to the 
direction of light propagation. This approach simplifies controlling both 
$\varphi$ and the total field as the angle or magnitude of $\vec{B}_0$ is 
varied.

\subsection{Measurements}
\begin{figure}[ht]
\includegraphics[width=8.0cm]{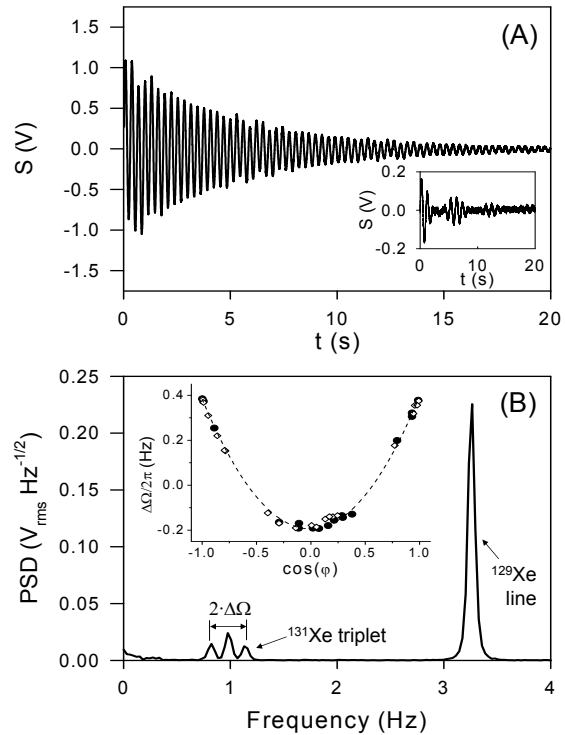}
\caption{\label{SampleData} Examples of experimental data. (A). A 
free-induction decay. The signal is proportional to the amplitude of the 
modulation of the photodiode signal caused by the Rb precession and is 
presented in volts. The magnetic field during the probe phase was 
$275 \,$nT, and $\varphi$ was $\sim 70 ^{\rm{o}}$. Beating from the $^{131}$Xe 
triplet is seen in inset, which shows the residuals from a fit of a damped 
sine wave to the data. (B). The Fourier transform of the data in (A). Here the 
signal is presented as a power spectral density in the units of root-mean-squared
volts in a 1 Hz bandwidth. The 
inset is a plot of the NQR splitting versus $\cos{\varphi}$. The solid symbols 
are data collected at $300 \,$mW/cm$^2$, and the open symbols are data 
collected at $60 \,$mW/cm$^2$. For the low (high) power data, the Rb field was 
0.12 $\mu$T (0.19 $\mu$T). For comparison, the applied field was 
$\sim 0.8 \mu$T. There is more distortion in the parabolic shape of the curve 
for the low power data -- particularly when the applied field was orthogonal 
to the light propagation direction where the approximation that 
$B_{Rb} \parallel \hat{k}$  would not hold as well. The dashed line is a fit 
to the high-power data, which gives a value of 
$\Delta\Omega_0/2\pi= 0.39(1)\,\rm{Hz}$. Our value for   $\Delta\Omega_0$ is 
$\sim 43$\,\% larger than the largest value reported by Wu et al. 
\cite{Wu1990}.}
\end{figure}

We concentrate our measurements on two things:  first, the $^{131}$Xe NQR shift 
$\Delta\Omega$ versus angle $\varphi$, and second, the transformation of the 
energy shifts as $B_0$ is swept from zero through the NQR-dominated regime and 
into the NMR-dominated regime. To measure $\Delta\Omega$ (Eq. 1) versus $\varphi$, 
$B_0$ was kept near 0.8$\,\mu$T. At high enough field, the NQR splitting 
depends on the field angle but not on the field magnitude. For this measurement, 
we did not apply a compensation field as described above, but rather included 
$\vec{B}_{Rb}$ in our calculation of 
$\varphi$. The data were consistent with $\vec{B}_{Rb} \parallel \hat{k}$. 
We measured the $^{129}$Xe and $^{131}$Xe spectra versus angle at two 
different laser powers: $0.6\,$mW and $3\,$mW (60 and 300 mW/cm$^{2}$). We 
determined frequency differences between the outer two $^{131}$Xe resonances 
using curve fitting and took $\Delta\Omega$ to be half of the difference. The 
inset in Fig. \ref{SampleData}B is a plot of  $\Delta\Omega$  versus 
$\cos\varphi$  for two different laser powers. Only field angle values where 
the triplet is resolved for curve fitting are included. The parabolic fit of 
the data agrees well with $\Delta\Omega = 0$ at the magic angle of 
$\varphi = 54.7^{\rm{o}}$ ($\cos \varphi  = 0.578$). 

To make quantitative estimates of the mean rotation angle per wall collision, 
we apply Eq. \ref{Eq1} to our cubic cell. Assuming $\langle\theta\rangle$ for 
pyrex can be expressed in terms of $\langle\theta\rangle$ for silicon as 
$\left\langle\theta_p\right\rangle = \left\langle\theta_s\right\rangle+\delta$ 
one finds $\Delta\Omega = \Delta\Omega_0 \rm{P}_2\left(\cos\varphi\right)$, 
where $\Delta\Omega_0$ reduces to $v\delta/L$. $L$ is the length of a side of 
the cube, and $v$ is the mean thermal velocity. Fitting this expression to the 
data in the inset of Fig. \ref{SampleData}B, we find 
$\Delta\Omega_0 = 2 \pi \times 0.39\,$Hz. With $L = 1\,$mm, and 
$v$ = 281 m/s, we find $\delta = 8.7\,\mu$rad. Using Wu et al.'s measurement 
of $\langle \theta_p \rangle = 38\,\mu$rad for pyrex \cite{Wu1990}, we find  
$\langle \theta_s \rangle = 29\,\mu$rad for silicon.

\begin{figure}[ht]
\includegraphics[width=7.0cm]{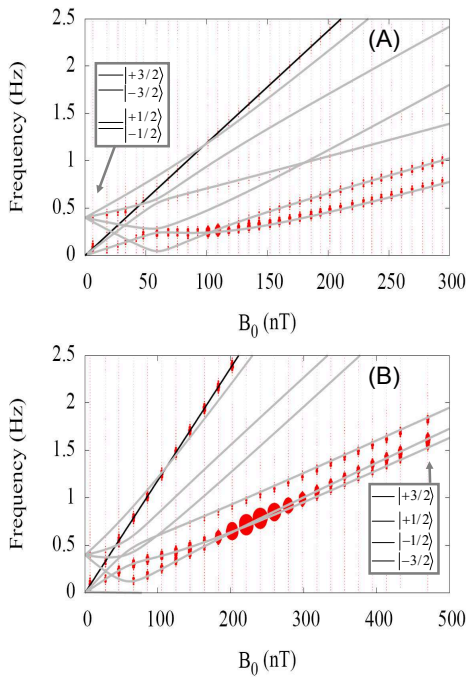}
\caption{\label{SpectraVsField} Spectra versus applied field for 
$\varphi = 22^{\rm{o}}$(A) and $\varphi = 39^{\rm{o}}$(B). 
The grey lines represent the six frequency differences between the four nuclear 
sublevels computed numerically (see text). The black lines mark the $^{129}$Xe 
transition versus field. Each vertical line 
represents a different frequency spectrum acquired at the specific magnetic 
field that it intersects on the horizontal axis. The symbol size is 
proportional to the signal amplitude, and the same amplitude scale was used 
for both angles of $\varphi$. The laser intensity was $300 \,$mW/cm$^2$ and 
$B_{Rb}$ was $0.19 \,\mu$T. The inset in (A) shows the $^{131}$Xe nuclear 
energy levels in the low-field NQR regime. Since the energies of the nuclear 
sublevels go as $m^2$ at zero magnetic field \cite{Pound1950}, the 
$\left|+ 3/2 \right\rangle$ and $\left|- 3/2 \right\rangle$ states are 
degenerate at zero field, as are the  $\left|+ 1/2 \right\rangle$ and 
$\left|- 1/2 \right\rangle$ states. The inset in (B) shows the nuclear energy 
levels in the high-field NMR regime.}
\end{figure}

To measure the spectra versus field magnitude for a fixed field angle, we 
carefully offset $\vec{B}_{Rb}$ by applying a compensation field along 
$\hat{k}$. We collected multiple spectra versus field magnitude at several 
field angles. Data are presented in Fig. \ref{SpectraVsField} for 
$\varphi = $22$^{\rm{o}}$ and 39$^{\rm{o}}$. The plots are three-dimensional --
the vertical axis is the measured frequency, the horizontal axis is the 
applied magnetic field (not counting $B_{comp}$), and the symbol size is 
proportional to the signal amplitude. Each vertical line represents an 
individual frequency spectrum collected at a fixed magnetic field. 
The black lines show the transition frequencies for $^{129}$Xe versus magnetic 
field. Since $^{129}$Xe has no nuclear electric quadrupole moment, the 
transition energy is linear in applied field. The gray lines represent the energy 
differences between the four nuclear sublevels of $^{131}$Xe found 
numerically, as discussed below.

\section{Calculations}
In the limit that the NQR shift is either much smaller or much larger than the 
Larmor frequency, perturbation theory solutions accurately predict the transition 
frequencies \cite{Bloom1955, Volk1979}.
When the NMR and NQR interactions are of comparable size,  a more involved solution 
is required. It is possible to solve the system analytically by 
diagonalizing the full Hamiltonian to find the transition energies as well as the 
transition amplitudes \cite{RochesterPrivateCommunication}, but it is involved -- 
particularly given the dynamics and the 
angular sensitivity of the Rb magnetometer. Finding the full analytical solution is 
made dramatically more accessible by taking as a starting point a general solution 
that has already been developed and is available online \cite{RochesterMathematica}. 
A full calculation yielding predictions for the line amplitudes that will give more 
insight into the physics is underway and will 
be reported elsewhere. For this work, we compare our data to 
numerical calculations using a Liouvillian approach described in detail by 
Bain \cite{Bain2003}. The method is relatively straightforward to use for calculating 
the transition frequencies for an arbitrary spin nucleus in an arbitrary 
magnetic field and electric field gradient, but it does not predict 
the transition amplitudes and it is also not very transparent. The details of the 
calculation are outside of the scope of this article, and we refer the interested 
reader to the original work \cite{Bain2003}, which gives the recipe for the 
calculations.
 
Three parameters enter the calculation: the angle $\varphi$, $\Delta\Omega_0$,
and an asymmetry parameter, $\eta$, which is zero in the case of a cylindrically 
symmetric electric field gradient. We set $\eta = 0$ for our simulations and for 
$\Delta\Omega_0$ we use our measurement from the inset of Fig. \ref{SampleData}B. We  
used our estimates of $\varphi$ from our coil calibrations and cell 
orientation assuming the symmetry axis of the electric field gradient to be along
the cell symmetry axis. We conservatively estimate an upper limit of 
$5^{\rm{o}}$ for angular misalignment of the cell axis from the magnetic field 
axis. 
 
The gray lines presented in Fig. \ref{SpectraVsField} are the six transition 
frequencies between the four nuclear states. 
At high magnetic fields in the NMR regime, there are three lines with equal 
slope, corresponding to $\Delta m=1$ transitions 
($\left|-3/2\rangle\langle-1/2\right|$, $\left|-1/2\rangle\langle1/2\right|$, 
and $\left|1/2\rangle\langle3/2\right|$), two lines with double the slope for 
$\Delta m=2$ transitions ($\left|-3/2\rangle\langle1/2\right|$ and 
$\left|-1/2\rangle\langle3/2\right|$), and one line for the $\Delta m=3$ 
transition ($\left|-3/2\rangle\langle3/2\right|$). 
Note that these transition frequencies are plotted with no insight into the 
transition amplitudes, and in fact, only the $\Delta m = 1$ transitions are 
allowed at higher fields. At very low magnetic field, $\Delta m =$ 1, 2, and 3 
transitions are also seen, but they do not correspond to the same $\Delta m =$ 
1, 2, and 3 transitions seen at higher field because the quantization axis 
rotates as a function of magnetic field thus transforming the states.

\begin{figure}[b]
\includegraphics[width=7.0cm]{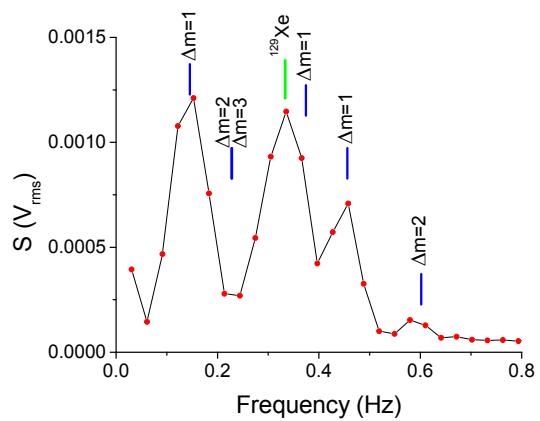}
\caption{\label{SingleSpectrum} A single spectrum collected with $28 \,$nT and 
$\varphi = 39^{\rm{o}}$, which corresponds to the second spectrum from the 
left in Fig. \ref{SpectraVsField}. The predicted transition frequencies are 
also shown for reference. The signal has the units of root-mean-squared amplitude 
for a 32.768 s averaging time. Given that the $T_2$ time is about 5 seconds, we 
compromised on signal amplitude to achieve improved frequency resolution.}
\end{figure}

The observed $^{131}$Xe lines agree well with the predicted frequencies even 
as the magnetic field goes to zero.
At low field, the transitions are unresolved, and in most cases it is 
difficult to assign features to $\Delta m=2$  or $\Delta m=3$ transitions. 
Perhaps the best resolved spectrum at low field, where we would expect strong 
mixing of the lines, was collected for $28\,$nT and $\varphi =  39^{\rm{o}}$, 
which corresponds to the second spectrum from the left in Fig. 
\ref{SpectraVsField}B. This spectrum is shown in Fig. \ref{SingleSpectrum}. 
The predicted transition frequencies are marked. One of the $\Delta m=2$ 
transitions is clearly visible and is one-tenth as strong as the $\Delta m=1$ 
transitions. The other $\Delta m=2$ transition and the $\Delta m=3$ transition 
are not visible. We cannot put limits on the strength of the $\Delta m=3$ 
transition, since the frequency where it would appear is obscured by the 
wings of other, much stronger transitions.

\section{Discussion and Outlook}
One point requiring further investigation relates to the amplitudes of the 
$^{131}$Xe lines. Whereas the $^{129}$Xe line amplitudes vary by about 10\% as 
$B_0$ is varied, the $^{131}$Xe line amplitudes vary by much more -- sometimes 
jumping up by a factor of 2 to 3 when the lines cross, and sometimes fading 
away and disappearing at a different field. The variations in line amplitudes 
for $^{131}$Xe will be the subject of further study in conjuction with 
performing a full analytical solution.

Our NQR shifts are large enough that we are able to measure the decay rates for 
the individual lines. Like the line amplitudes, the decay rates for the $^{131}$Xe 
lines vary widely -- especially at the line crossings. The ratio of the decay 
rates for the $\left|-3/2\rangle\langle-1/2\right|$ and 
and $\left|1/2\rangle\langle3/2\right|$ coherences relative to the 
$\left|-1/2\rangle\langle1/2\right|$ coherence does not agree with the 3:2 
result predicted by Wu et al. for the NMR regime \cite{Wu1988}. The variation in 
the decay rates will also be an area of further study.
 
\section*{ACKNOWLEDGMENTS}
We thank Michael Romalis for initially suggesting that we study this very 
interesting regime, Simon Rochester and Dmitry Budker for assistence with 
and insight into the theoretical calculations, and Will Happer, Hugh Robinson, 
Svenja Knappe, Susan 
Schima, and Leo Hollberg for technical assistance and discussions. This work 
was funded by DARPA and is a contribution of NIST, an agency of the US 
government, and is not subject to copyright.
\end{document}